\documentclass []{aa}
\usepackage{psfig} 
\begin{document}
\thesaurus{03.13.4, 03.13.6, 11.03.1}
\title{Uncertainties on Clusters of Galaxies Distances.}

\author{C.~Adami \inst{1,2}, M.P.~Ulmer \inst{2}}

\institute{
LAM, Traverse du Siphon, F-13012 Marseille, France  
\and Department of Physics and Astronomy, NU, Dearborn
Observatory, 2131 Sheridan, 60208-2900 Evanston, USA 
}
\offprints{C.~Adami} 
\date{Received date; accepted date} 

\maketitle 
 
\markboth{Uncertainties on Clusters of Galaxies Distances.}{} 
 
\begin{abstract} 
 
We investigate in this paper the error on the cluster redshift estimate as 
a function of: (1) the
number of galaxy structures along the line of sight; (2) the morphology of
the clusters (regular/substructures); (3) the nature of the observed galaxies
(cD/normal galaxies); (4) the number of observed galaxies; and (5) the 
distance of the clusters. We find that if we use cD 
galaxies we have errors of less than 2\%  (at the 1-$\sigma$ level) for the 
cluster distances except when the clusters are very complex. 
In those cases when we use five redshift 
measurements along the cluster lines of sight to compute the mean redshift, the 
mean redshift estimate error is less than 5$\%$ for the clusters closer than 
z=0.30. The same error of 5\% for clusters in z=[0.30;0.60] requires about 12 
redshifts measurements. 

\end{abstract} 
 
\begin{keywords} 
 
{ 
Methods: numerical ; Methods: statistical ; Galaxies: clusters: general
} 
 
\end{keywords} 

\section{Introduction}

Large surveys of clusters of galaxies are commonly used in modern
observational cosmology both for optical (e.g. ENACS: Katgert et
al. 1996, CNOC: Carlberg et al. 1996) and X-ray studies (e.g.  Ebeling
et al. 1998, Rosati et al. 1998, Vikhlinin et al. 1998, Romer et al.
2000).  Only the mean redshift of the clusters is necessary for example
for constraining cosmological models using the N/z relation (Oukbir $\&$ 
Blanchard 1997), i.e. the variation of the cluster abundance as a 
function of redshift and mass. Since masses of these
clusters can be measured from X-ray data, thus spectroscopy is only
necessary to compute the mean redshift. For very large
(typically more than 100 clusters) and deep surveys, however, it is
difficult to measure a large number of redshifts for each cluster due to
observing time constraints. But, too low a number of redshifts can induce
errors for the mean redshift. We propose in this work to quantify 
these errors by using simulations. The purpose of this paper is to 
provide guidelines for estimating statistical errors on the 
distance of clusters of galaxies in existing catalogs as a function 
of the galaxies sampled (number of redshifts per line of sight
and nature of the galaxies) and of the distances and
characteristics of these clusters. Estimates  based only on a small 
number of redshifts and a limited number of cases can be found in Holden 
et al. (2000).

The first part of the article describes the method and the sample.
The second part gives the results of the simulations. 
The third part discusses these results. 

We used H$_0$=100 km.s$^{-1}$.Mpc$^{-1}$ and q$_0$=0.05.

\section{General methodology and Sample.}

We have tested the influence on the mean redshift uncertainty of: (1) the
number of galaxy structures along the line of sight; (2) the morphology of
the clusters (regular/substructures); (3) the nature of the observed galaxies
(cD/normal galaxies); (4) the number of observed galaxies; and (5) the 
distance of the clusters.                                         
                                        
These clusters were selected from the ENACS database
(Katgert et al. 1996), from a literature compilation by A. Biviano
(described for example in Adami et al. 1998), and from the public CNOC
database (Carlberg et al. 1996). We only selected the very well
sampled clusters, i.e. those with a sufficient number of redshifts to
allow us to assume safely that the redshift measurements provided the
true value of the mean cluster redshift. We then resampled each of
these clusters with $n$ galaxies before re-computing the mean redshift
to estimate the error for this redshift. The method we used to compute 
the mean redshift
after the resampling with $n$ galaxies is as follows: we used the same
analysis methods as for the ENACS lines of sight (Katgert et
al. 1996). We excluded the galaxies classified as field objects (see
below) and we used a Biweight estimator (Beers et al. 1990) of the
mean for the cluster galaxies. Briefly, field galaxies were
discriminated by searching for velocity gaps of more than 1000
km.s$^{-1}$ in the redshift distributions of the galaxies sorted by
increasing redshift. If two galaxies were separated by more than 1000
km.s$^{-1}$, they were assumed to belong to different structures. The
method is explained in detail in Katgert et al. (1996), but we note
that the choice of a value of 1000 km.s$^{-1}$ for the
gap has no significant influence on the structure membership determination 
(see Adami et al. 1998): a value between $\sim$600 and $\sim$1200 km.s$^{-1}$
gives similar results.

In order to remove possible interloper clusters (with too small a
number of redshifts) from the sample, we selected only the lines of
sight with more than 50 redshifts in the main groups (the ``clusters''
hereafter). We also discarded the lines of sight with too high a
contamination level in a first step: we rejected the lines of sight
with more than two structures sampled with more than 20
redshifts. This limitation corresponds approximatively to the separation
between the massive and minor groups in the ENACS sample.
These lines of sight are considered in section 3.3.  At
this step, there were 73 lines of sight in the sample. As we used a
literature compilation, selection effects are sometimes
significant. For example, galaxies along a few lines of sight in
the sample were observed by using a color-magnitude selection of the
targets (Mazure et al. 1988) or only the cluster galaxy redshifts were
given in the literature. This potentially induced a bias toward the
cluster galaxies. In order to limit this problem, we discarded all the
lines of sight with more than 95\% of the galaxies inside the main
cluster (cf. Fig. 1) because these lines of sight were very likely to be
biased toward the cluster galaxies. This value of 95\% is based on our 
analysis of the ENACS results (Katgert et al. 1996). In this survey there 
was no selection of galaxies with a color-magnitude relation. All the
redshifts measured along the lines of sight were provided. We found that these
ENACS lines did not have more than 95\% of the galaxies within a single
structure.

In total, after all the rejections, we used a sample of 58 lines of
sight (Table 1).
We split this sample of 58 clusters in five different redshift bins in
order to compute the redshift estimate precision as a function of the
cluster distance (and of the number $n$ of sampling galaxies). The first
bin (z=[0; 0.07]) contains 37 clusters, the second bin (z=[0.07; 0.13])
contains 14 clusters, the third and the fourth bins (z=[0.13,0.30] and
z=[0.30,0.45]) contain 3 clusters and the last bin (z=[0.45,0.60]) contains
2 clusters. We note that the fourth and the last bins mainly used CNOC 
data (Carlberg et al. 1996).

\section{The results}

\subsection{Lines of sight with only one dominant structure}

\begin{figure} 
\vbox 
{\psfig{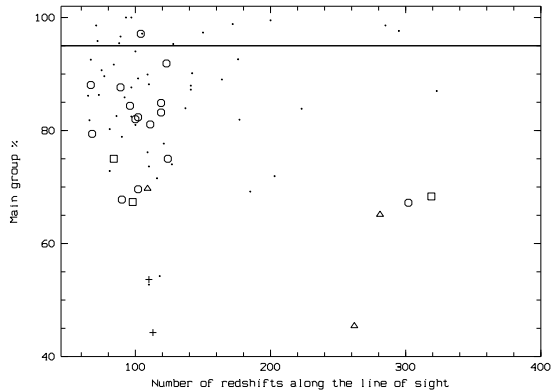}} 
\caption[]{Percentage of galaxies belonging to clusters along the 73 
lines of sight (i.e. before rejection of the highly contaminated
lines). Table 1 gives only the 58 lines of sight with a percentage
lower than 95\%. Small dots: clusters in the first redshift bin,
circles: clusters in the second redshift bin, squares and 
triangles: clusters in the third and in the fourth redshift
bins, crosses: the two clusters in the most distant redshift bin
} 
\label{} 
\end{figure}

We resampled all the 58 lines of sight (i.e. with only one dominant
structure) with $n$ galaxies, with n=[5;47] (the method
we used to compute the mean redshift is not effective for a sampling rate
of less than 5 galaxies). In order to have the same statistical
significance for each of the five redshift bins (they do not have the
same number of clusters), we made about 5000 realizations (bootstrap
technique with 5000 resampling) for a given
$n$ (number of selected redshifts along the line of sight) and for a
given cluster redshift range. This means that we made 135 different
resamplings for each line of sight and each $n$ in the first redshift
bin (37 clusters $\times$ 135 simulations $\sim$5000 simulations), 357
different resamplings for each line of sight and each $n$ in the
second redshift bin, 1667 different resamplings for each line of sight
and each $n$ in the third and fourth redshift bins, and 2500 different
resamplings for each line of sight and each $n$ in the fifth redshift
bin.

\begin{table*} 
\caption[]{col 1: cluster name; col. 2: z bin; col 3: number of gal. along the
$los$; col. 4: gal. in the clusters; col 5: cluster velocities: km.s$^{-1}$;
col 6: distance between the X-ray center and the 1st ranked galaxy; col 7: 
difference between the mean velocity of the cluster and the velocity of the 
first ranked galaxy; col8: S: substructured clusters, R: regular clusters.}
\begin{flushleft} 
\begin{tabular}{cccccccc} 
\noalign{\smallskip} 
name & bin & $los$ galaxies & cluster galaxies & cz & distance (kpc) & vel. 
difference (km.s$^{-1}$) & substructures \\ 
\hline 
\noalign{\smallskip} 
A85	& 1	& 185	& 128	& 16384 & & & S \\
A119	& 1	& 142	& 128	& 12958 & 15 & 63 & R \\
A168	& 1	& 109	& 83	& 13216 & 106 & 51 & S \\
A193	& 1	& 65	& 56	& 14285 & 22 & 125 & R \\
A400	& 1	& 109	& 98	& 6952 & 4 & 87 & \\
A496	& 1	& 164	& 146	& 9768 & 18 & 25 & R \\
A754	& 1	& 100	& 94	& 16635 & 493 & 110 & R \\
A978	& 1	& 73	& 63	& 16689 & & & S \\
A1060	& 1	& 177	& 145	& 4000 & 19 & 54 & S \\
A1185	& 1	& 77	& 69	& 9704 & 92 & 1113 & \\
A1631	& 1	& 90	& 71	& 14243 & 354 & 283 & R \\
A1644	& 1	& 102	& 91	& 14504 & 33 & 55 & S \\
A1795	& 1	& 97	& 85	& 19065 & 31 & 14 & R \\
A1983	& 1	& 100	& 81	& 13589 & 1102 & 274 & S \\
A2040	& 1	& 66	& 54	& 13976 & 24 & 133 & R \\
A2052	& 1	& 97	& 80	& 10636 & 8 & 150 & S \\
A2107	& 1	& 75	& 68	& 12474 & 7 & 265 & S \\
A2124	& 1	& 67	& 62	& 19589 & & & R \\
A2717	& 1	& 81	& 59	& 14408 & & & R \\
A2734	& 1	& 116	& 83	& 18121 & & & R \\
A2877	& 1	& 110	& 97	& 7037 & 23 & 7 & \\
A3122	& 1	& 121	& 94	& 19173 & & & S \\
A3128	& 1	& 223	& 187	& 17927 & 279 & 412 & S \\
A3158	& 1	& 141	& 123	& 17606 & 43 & 213 & R \\
A3223	& 1	& 110	& 81	& 18035 & 79 & 1158 & S \\
A3341	& 1	& 118	& 64	& 11377 & & & S \\
A3354	& 1	& 110	& 58	& 17604 & & & S \\
A3376	& 1	& 84	& 77	& 14063 & & & \\
A3391	& 1	& 81	& 65	& 16302 & & & \\
A3395	& 1	& 203	& 146	& 15283 & & & \\
A3558	& 1	& 323	& 281	& 14612 & 5 & 270 & R \\
A3651	& 1	& 92	& 79	& 17817 & & & S \\
A3667	& 1	& 176	& 163	& 16557 & 1 & 73 & R \\
A3716	& 1	& 137	& 115	& 13792 & & & \\
A3744	& 1	& 86	& 71	& 11169 & & & S \\
A3809	& 1	& 127	& 94	& 18430 & & & R \\
A401	& 2	& 123	& 113	& 21868 & 92 & 591 & R \\
A514	& 2	& 111	& 90	& 21589 & 778 & 892 & R \\
A1809	& 2	& 67	& 59	& 23985 & 13 & 69 & R \\
A2029	& 2	& 89	& 78	& 23124 & 9 & 458 & \\
A2142	& 2	& 119	& 101	& 27107 & & & \\
A2670	& 2	& 302	& 203	& 22494 & 15 & 447 & S \\
A2721	& 2	& 100	& 82	& 34039 & & & \\
A3094	& 2	& 102	& 71	& 20004 & & & S \\
A3112	& 2	& 124	& 93	& 22460 & 4 & 203 & R \\
A3695	& 2	& 96	& 81	& 26678 & & & S \\
A3705	& 2	& 68	& 54	& 26647 & & & S \\
A3806	& 2	& 119	& 99	& 22772 & & & S \\
A3822	& 2	& 102	& 84	& 22536 & & & S \\
A3825	& 2	& 90	& 61	& 22327 & 386 & 116 & R \\
A1146	& 3	& 84	& 63	& 43207 & & & \\
A2390 	& 3	& 319	& 218	& 68370 & & & \\
A3888	& 3	& 98	& 66	& 45434 & & & \\
MS1008-12 & 4	& 109	& 76	& 91860 & 87 & 63 & R \\
MS1358+62 & 4	& 281	& 183	& 98700 & 202 & 780 & S \\
MS1621+26 & 4	& 262	& 119	& 128220 & & & S \\
MS0015+16 & 5	& 110	& 59	& 162270 & & & R \\
MS0451-03 & 5	& 113	& 50	& 158790 & & & R \\
\noalign{\smallskip} 
\hline	    
\normalsize 
\end{tabular} 
\end{flushleft} 
\label{} 
\end{table*}

For each of the 5000 realizations we computed the difference between the true 
value of the cluster redshift and the estimated value of this redshift. We 
finally computed the dispersion of the 5000 percentage differences. The mean 
values of these difference percentages were
always consistent with an error of 0\% and in the range $\pm 2\%$. We
focus here on the variation of the dispersion $\sigma $ of these 5000
differences as a function $n$
(the number of selected redshifts along the line of sight). These variations 
are given in Fig. 2 for the five different cluster redshift bins. 
The
1-$\sigma $ error is always less than 5\% for the clusters at z$\leq$0.30, 
whatever the number of redshifts (greater than 5) we use along the
line of sight to compute the mean distances of the clusters. These errors
become greater than 5\% (but are still lower than 10\%) for the clusters
more distant than z=0.30 when we use less than 13 galaxies along the lines
of sight to compute the mean redshifts of the clusters.

\subsection{Influence of the nature of the clusters: regular vs
substructured clusters.}

The dynamical state of the clusters is also important for the redshift 
estimate: regular or with more perturbed history.
When the clusters have a very complex velocity structure we need to know if 
the statistical approach
of using a larger number of fainter galaxies will succeed.

\begin{figure} 
\vbox 
{\psfig{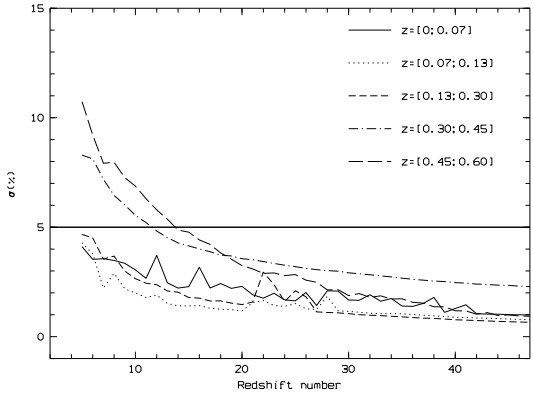}} 
\caption[]{1-$\sigma$ redshift error in percentage for the clusters of the
bins 1 to 5 of
Section 3.1, using all the clusters. The horizontal thick line
is the 5$\%$ level.} 
\label{} 
\end{figure}

We have chosen to repeat the analysis of Section 3.1.
This time we used a discrimination between the regular and the irregular 
clusters. We used the work of Bird (1994), Bird $\&$ Beers (1993) and Solanes 
et al. (1999) to produce a classification for the Abell clusters in our sample. 
These authors have performed substructure tests with a subsample of the 
clusters in Table 1. We classified a cluster of Table 1 as regular only if no 
one of the tests provided by these authors gave evidence of substructures at 
the 10$\%$ level. For the CNOC clusters, we plotted galaxy isodensity contours 
and we selected as regular the clusters providing regular contours. We assumed 
that no substructures in a cluster was an indication of a quiescent history.
The same kind of classification could possibly be achieved in the future 
using X-ray data. However, for most of the clusters in our sample, the only 
X-ray images available have too low a spatial resolution or too weak a signal
to allow us to resolve substructures for the distant clusters.

We have separated the clusters of the sample with substructure information
into 3 redshift bins: the bins 1, 2 and 5 of Section 3.1 (the bins 3 and 4 do 
not provide alone robust estimates). We plot on Fig. 3 the 
variation of the redshift error percentage as a function of the number of 
measured galaxies used to compute the mean distance of the cluster for the 
regular clusters. This is directly comparable with the percentages given in 
Fig. 2. We see that the error percentages are slightly lower when we use 
the regular clusters compared to the whole sample (regular + substructured + 
unknown). The difference is close to 1 or 2 percent. This means that the 
clusters with substructures do not produce very significantly higher errors in 
the mean redshift estimate. The approach of Section 3.1 is, therefore, not very 
sensitive to the nature of the clusters. 

\subsection{Lines of sight with more than one dominant structure}

\begin{figure} 
\vbox 
{\psfig{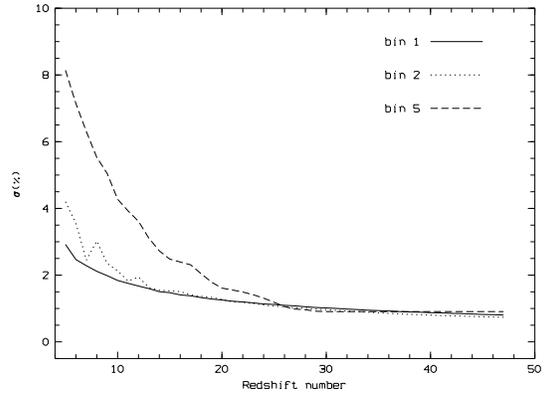}} 
\caption[]{1-$\sigma$ redshift error in percentage for the clusters of the
bins 1 (lower curve), 2 (middle curve) and 3 (higher curve) of
Section 3.1 and Fig. 2, using only the regular clusters.} 
\label{} 
\end{figure}

We only use here the clusters satisfying the conditions of Sect. 3.1,
but with at least one group with more than 20 galaxies in addition to
the main structure. These clusters are A151 (z=0.052, 3 groups with
more than 20 gal.), A1656 (z=0.024, 3 groups with more than 20 gal.),
A2390 (z=0.23, 2 groups with more than 20 gal.), A2634 (z=0.030, 2
groups with more than 20 gal.), A4038 (z=0.029, 5 groups with more
than 20 gal.) and MS1512+36 (z=0.37, 3 groups with more than 20
gal.). The results of the simulations show a higher 1-$\sigma$ error
in percentage, as expected, but the difference is not very
significant. First, we have no difference when we use more than 10
redshifts, whatever the distance of the cluster is. When we used less than
10 redshifts, we add about 2$\%$ to the values of Fig. 2.  We have an
unchanged number of redshifts required to have an error smaller than
5$\%$ for z$\leq$0.3. For z$\geq$0.3, we now need more than 8 or 9
redshifts measurements (instead of 5). We also note that such
superpositions are not very likely. They represent only $\sim$10$\%$
of the total sample we used in this study. We will, therefore, refer
only to the Fig. 2 results.

\subsection{Influence of the nature of the galaxies: using the 
cD velocities.}

The three previous sub-sections deal with normal galaxies. However, an
important question with fundamental implications is to know if the first
ranked galaxy (assumed to be the central Dominant galaxy hereafter) is
a good measure of the cluster redshift. This kind of galaxy is, in a simple
picture, at the bottom of the cluster potential, at the spatial center of the 
cluster and, therefore, at a redshift very close to the mean cluster redshift. 
In theory, this allows the cluster redshift measurement to be made with only 
one galaxy (e.g. Gunn et al. 1986, Hill $\&$ Oegerle 1998). However, when the 
cluster history is complex (mergings), the cD galaxies can be sometimes 
classified as $speeding~cD's$ (see e.g. Zabludoff et al. 1993 and references 
therein), i.e. cD galaxies with a significant velocity difference compared to 
the mean cluster velocity. Moreover, those can be shifted compared to the 
cluster center (e.g. Peres et al. 1998). In this case, the use of the cD
alone can fail in measuring accurately the cluster mean redshift. 

In order to investigate this point, we have used the clusters of our real 
sample using the first ranked galaxies with a measured redshift. Moreover, to 
separate the normal cD's and the speeding cD's, we used only the clusters 
with an X-ray center (see Tab. 1). The peak emission of the X-ray gas is 
assumed to be the real center. The distance between the cD and the X-ray 
center is used as an indication of the nature of the cD galaxies. When the
cD galaxy  is located at less than 50 kpc in projected distance from the 
cluster center (the X-ray center), the difference between the cluster mean 
velocity and the cD velocity is (0.15$\pm$0.12)$\times$$\sigma _v$ (with 
$\sigma _v$ the biweight velocity dispersion of the cluster). For the
cD's at more than 50 kpc, the difference is (0.53$\pm$0.47)$\times$
$\sigma _v$. Despite the large uncertainties, it seems clear that when
the cD's are far from the X-ray center, they provide a less reliable
estimate of the mean cluster velocity: for a typical cluster with a
velocity dispersion of 1000 km.s$^{-1}$, the mean difference is
$\sim$550 km.s$^{-1}$ ($\sim$150 km.s$^{-1}$ when X-ray center and cD
position match). 

Besides testing the cD derived redshift as a function of the distance from the
cluster center, we also tested the cD versus the mean redshift (as identify
the true cD can be more difficult with increasing redshift). In order
to have enough clusters with an X-ray center in each bin, we could only
separate the sample into two bins: redshifts lower and higher than
0.06. For the nearby clusters and the cD galaxies close to the center,
the error on the cluster mean redshift estimate is 
(0.14$\pm$0.11)$\times$$\sigma _v$. For the 
nearby clusters and the cD galaxies far from the center, the error 
is (0.39$\pm$0.29)$\times$$\sigma _v$. For the more distant clusters 
and the cD galaxies close to the center, the difference is (0.19$\pm$0.16)
$\times$$\sigma _v$. Finally, for the more distant clusters and the cD 
galaxies far from the center, the difference is (0.68$\pm$0.60)$\times$
$\sigma _v$. According to the uncertainties, the 
differences are not very significant but still very suggestive. Whatever 
the cluster redshift (nearby or more distant), the cD galaxies close to the 
center are better estimators of the mean cluster velocity compared to the cD 
galaxies far from the center. However, this trend tends to become stronger 
at higher redshifts, perhaps due to the fact that we are dealing with younger
structures, therefore at earlier formation stage and smaller than present day
clusters.

The conclusion of this section is that the cD galaxies are good estimators
of the cluster velocities when they are not speeding cD galaxies. Given
in the same units as in the previous sections, the redshift error is lower
than 1$\%$ when we use the cD galaxy redshift as the cluster redshift.

When we deal with disturbed clusters and speeding cD galaxies, the 
error is larger but often acceptable: for example for clusters at
z=0.05 and with a velocity dispersion of 1000 km.s$^{-1}$, the error on
the mean redshift estimate is 3$\pm$2 $\%$. However, in a few cases, 
this error can be close to 10$\%$ (at the 3-$\sigma$ level). Moreover, for 
the very distant cluster, it is not trivial to find which galaxy is the 
dominant one due to foreground galaxies. In these cases, the cD 
approach fails to measure the mean distance of the clusters.

\section{Summary}

We have used a sample of 58 rich clusters of galaxies (from z=0 to 
z=0.60) in order to investigate the error 
on the cluster redshift estimate as a function of the distance of the cluster, 
of the number $n$ of measured redshifts along the line of sight, of the 
nature of the 
galaxies (cD galaxies or normal galaxies) and of the nature of the clusters
(regular/substructures).
 
The mean redshift of the clusters are used, for example, to compute the 
absolute magnitudes of the galaxies in a given cluster from the observed
apparent magnitudes (luminosity function studies, e.g. Lumsden et
al. 1997, Valotto et al. 1997, Rauzy et al. 1998) or to measure the
X-ray luminosity of this cluster (e.g. Romer et al. 2000) from the
observed X-ray flux. If we require an error in the optical magnitude
of a cluster galaxy of less than 0.25 magnitude or of less than 10\%
on a cluster X-ray luminosity, we need to measure the mean redshift of
this cluster with an accuracy of less than $\sim $5\%. For nearby
clusters (z$\leq$0.30), if we use our method to compute the mean
redshift (discrimination of field galaxies with a gap analysis and
Biweight estimators of the mean: Beers et al. 1990), the error on the
redshift estimate is better than 5\% even with only 5 redshift
measurements along the line of sight.

If we now consider more distant clusters (z in the [0.30;0.60] range), we have
an error of less than 5\% if we measure more than 12 redshifts along
the line of sight. With only 5 redshifts, the error on an optical
magnitude can be (at the 67\% level) 0.5 mag and that on an X-ray
luminosity will be 20\%. These
estimates use galaxy redshifts distributed in a region of about 4
h$^{-2}$Mpc$^2$ (the typical size covered for each cluster in clusters
of galaxy surveys such as ENACS: Katgert et al. 1996) around the
cluster centers and magnitude without selection criteria. This means that 
the errors we plotted in Fig. 2 probably would decrease with a more 
concentrated survey around the cluster centers (but we have not tested this
point in this article). 

We show that it is slightly easier to measure an accurate mean
redshift for the regular clusters (with quiescent history and, therefore, 
with symmetric potential well) using the approach of Section 3.1 (measuring a 
few dozens of redshifts). This method does not provide, however, significantly 
worse results for the substructured clusters (with more perturbed an history).

Finally, we showed that for the regular clusters, the measure of the
cD galaxy redshift is a good way to measure accurately the mean redshift of
the parent-cluster. Sometimes, however, this approach fails (for the 
substructured clusters or very distant 
clusters where the cD nature of a given galaxy is not straightforward to 
define). This approach should be replaced with a multi-redshift measure. The 
5$\%$ 
error level is reached with 5 redshift measurements for the cluster closer than 
z=0.30 and with about 12 redshift measurements for the more distant clusters 
(z$\leq$0.60). 

\begin{acknowledgements}

{CA thanks the Dearborn Obs. staff for their hospitality during his 
postdoctoral fellowship. CA and MPU thank the referees for useful and
constructive comments, Florence Durret for a careful reading of the 
manuscript and Andrea Biviano for his compilation. CA thanks the LAM for
support.}

\end{acknowledgements}

\vfill 
\end{document}